\documentclass[final,5p,times,twocolumn]{elsarticle}

\usepackage{graphicx}
\usepackage{amssymb}
\usepackage{epstopdf}
\usepackage{hyperref}

\journal{Physica C}

\begin{document}

\begin{frontmatter}

\title{Quantum criticality and the phase diagram of the cuprates}

\author{Subir Sachdev}

\address{Department of Physics, Harvard University, Cambridge MA 02138}
\address{E-mail: sachdev@physics.harvard.edu}

\begin{abstract}
I discuss a proposed phase diagram of the cuprate superconductors as a function of temperature,
carrier concentration, and a strong magnetic field perpendicular to the layers. I show how the phase
diagram gives a unified interpretation of a number of recent experiments.\\
~\\
{\em Keynote talk, 9th International Conference on Materials and Mechanisms of Superconductivity, Tokyo, Sep 7-12, 2009.}
\end{abstract}


\end{frontmatter}

\begin{figure}[h!]%
\includegraphics*[width=0.5\textwidth]{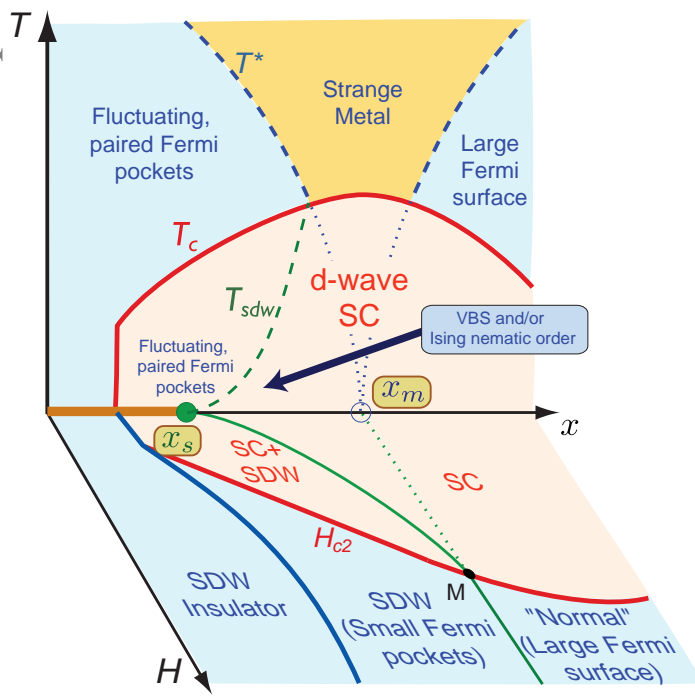}
\caption{Proposed phase diagram of the cuprates showing the interplay between
superconductivity (SC), spin density wave (SDW) order, and Fermi surface configuration
as a function of carrier density ($x$), temperature ($T$), and magnetic field ($H$) 
perpendicular to the layers. Full lines are thermal or quantum phase transitions, dashed lines are crossovers, 
and dotted lines are guides to the eye.
The phase transitions associated with valence bond solid (VBS) (or ``charge'') and nematic order are not shown.
The superconducting regions are colored pink. We have assumed the absence of interlayer coupling,
and so the SDW order is long-ranged only at $T=0$: it is present in the regions labeled ``SDW'' and on the thick orange
line for $x< x_s$. In the blue normal regions, the `pseudogap' is between $T_c$ and $T^\ast$, the `Strange Metal' has an in-plane
resistivity which is measured to be linear in $T$, and the ``Large Fermi surface'' has a conventional $T^2$ resistivity.}
\label{figglobal}
\end{figure}

\pagebreak
This brief note contains a summary of the key aspects of the proposed phase diagram of the cuprate
superconductors shown in Fig.~\ref{figglobal}, and the central role played by ideas of quantum 
criticality. 
A more detailed discussion can be found in another recent review by the author \cite{qcnp}, which also
contains more complete citations to the literature. Here, I will focus on the central physical ideas
and highlight support from recent experiments.

The phase transitions and crossovers in Fig.~\ref{figglobal} appear quite intricate. However, they
can be understood simply by focusing first on the quantum critical point (QCP) at doping density, $x=x_m$,
temperature $T=0$, magnetic field $H=0$. As indicated in Fig.~\ref{figglobal}, this quantum critical point
is pre-empted by the onset of superconductivity.

The QCP at $x=x_m$ is a transition between two metallic (hence the subscript $m$)
Fermi liquid phases. At $x>x_m$ we have the full symmetry of the square lattice, and a ``large'' Fermi surface metal
consisting of a hole-like Fermi surface enclosing the area $1+x$ expected from the Luttinger theorem (this is for hole doping;
with electron doping, $x$, the area enclosed is $1-x$). At $x<x_m$ we have the onset of spin density wave (SDW) order,
and this breaks apart the large Fermi surface into ``small'' Fermi pockets. Nevertheless the Luttinger theorem continues
to be obeyed, after accounting for the large unit cell created by the SDW order. The ultimate theory of this quantum
critical point is not fully understood, despite much theoretical attention \cite{acs}.

Strong evidence for the QCP at $x=x_m$, and its associated $T>0$ crossovers comes from recent experiments on 
Nd-LSCO \cite{nlsco1,nlsco4}. They detected the crossover between ``Strange Metal'' and ``Fluctuating paired Fermi pockets''
regions of Fig.~\ref{figglobal}. The latter region is our identification of the popular `pseudogap' phase \cite{gs}, and so the crossover
temperature is $T^\ast$. The experiments identified $T^\ast$ by deviations from linear resistivity in the in-plane resistance,
or by an upturn in the c-axis resistivity, and showed these were correlated with signatures of changes in the area enclosed
by the Fermi surface. Further, they were able to track crossovers down to $T=0$ by suppressing superconductivity by 
an applied magnetic field. This corresponds to locating the QCP along the green line beyond the point M
in the $T=0$ plane in Fig.~\ref{figglobal}.

Earlier evidence for the QCP at $x=x_m$ came indirectly from Panagopoulos and collaborators \cite{christos1,christos2}, who used 
 muon spin relaxation and ac-susceptibility measurements on a series of pure and Zn-substituted hole-doped cuprates
 to observe a glassy slowing down of spin fluctuations. This glassy behavior vanished above a critical doping which
 we identify as $x=x_m$.
 
Recent thermoelectric and Nernst effect experiments \cite{thermoelectric} and theory \cite{nernsttheory1,nernsttheory2}
have also provided support for the Fermi surface transformations associated with the QCP at $x=x_m$.
Associated measurements of the anisotropy in the Nernst co-efficient \cite{brs} have been proposed to be explained
by the influence of nematic order
in the Fermi surface \cite{nernsttheory3}; this nematic order can be regarded as a remnant of a fluctuating SDW state,
as is suggested by neutron scattering observations \cite{keimernematic}.

Finally, I also note the recent quantum oscillation observations \cite{helm} in the electron-doped cuprates, which show
striking direct evidence for the sudden change in Fermi surface area at a large $H$. We identify this as a signature
of the $T=0$ green transition line beyond the point M at $x=x_m$ in Fig.~\ref{figglobal}.
 
Now, let us consider the onset of superconductivity at $H=0$. This occurs in a dome-shaped region
around $x=x_m$ \cite{moon}. Here a crucial effect is that the competition between the SC
and SDW orders {\em shifts} the position of the SDW-ordering QCP to $x=x_s$ (the subscript
$s$ refers to the presence of superconductivity). Loosely speaking, the competition is for
the Fermi surface: both the SDW and SC orders want to induce gaps in the same regions of
the Fermi surface. We have presented a theory \cite{moon} which shows how
such a competition leads to the shift in the position of the QCP.

Evidence for the QCP at $x=x_s$ appeared in early neutron scattering studies \cite{keimerold},
and its signatures were addressed in initial theories for quantum criticality \cite{sy}. Also, in the electron-doped
cuprates, the value of $x_s$ has been quite precisely identified \cite{motoyama} --- the high field quantum oscillation
experiments we noted above \cite{helm} were done on the same material, and indeed found as $x_m  > x_s$.

With the shift in the QCP from $x_m$ to $x_s$, the crossovers as $T$ is lowered for $x_s < x < x_m$ at $H=0$
are quite complex, but simple to deduce from the topology of our phase diagram. 
As $T$ is reduced below $T^\ast$, the electrons start to develop signatures of the onset of local SDW (and associated nematic)
order. However, as $T$ approaches $T_c$, the competition with SC halts the march towards to stronger SDW ordering.
We sketch a crossover temperature $T_{\rm sdw}$ in Fig.~\ref{figglobal}, below which the electrons abandon SDW ordering,
and the physics of the underlying large Femi surface can reappear {\em i.e.} the spectrum of the Bogoliubov quasiparticle excitations
of the $d$-wave superconductor can lose signature of the Fermi pockets. Note that 
superconductivity competes mainly with SDW order,
and will have a weaker suppression effect on the associated tendencies to VBS/nematic ordering \cite{moon}. Indeed, these orderings can survive
at $T=0$, as has been discussed in some toy models \cite{rkk3}. Low $T$ evidence for such ordering, and their connection
to the pseudogap phase, has appeared in scanning tunnelling microscopy experiments \cite{kohsaka1,kohsaka2}.
 
Now let us consider phase diagram at $T=0$ in the $x$, $H$ plane. The general structure of the phase transitions here
appeared in early theoretic work \cite{demler}, and indeed motivated the $T>0$ portion of the phase diagram already discussed.
A key prediction of this work was that the shift in the QCP
from $x_m$ to $x_s$ implies the presence of a line of quantum phase transition within the SC
phase which connects the point $x_s$ to the point M in Fig.~\ref{figglobal}. This line marks the onset
of long-range SDW order. A number of recent experiments \cite{mesot,mesot2,mesot3} have presented strong evidence for
this transition, in both LSCO and YBCO.

Moving to stronger fields, we loose superconductivity at $H=H_{c2}$ and cross into the normal state.
The crucial, recent observation of high field quantum oscillations \cite{doiron,cooper,nigel,cyril,suchitra,louis,suchitra2,gil}
lead us to identify their small Fermi pockets with those of the normal phase region for $x<x_m$ in the $x$,$H$ plane.

Also shown in the $x$,$H$ plane of Fig.~\ref{figglobal} is a metal-insulator transition to a low-doping SDW insulator.
We believe this transition is associated with the localization of the small Fermi pockets, and is related to a number
of experimental observations \cite{greg,gil}.

\section*{Acknowledgements}

I would like to thank my experimental colleagues, G.~Boebinger, J.~C.~Davis, B.~Keimer, G.~Lonzarich, C.~Panagopoulos,
S.~Sebastian and L.~Taillefer,
for numerous enlightening discussions.
This research was supported by the NSF under grant DMR-0757145, by the FQXi
foundation, and by a MURI grant from AFOSR.

\end{document}